\documentclass[conference]{IEEEtran}
\IEEEoverridecommandlockouts
\usepackage{cite}
\usepackage{amsmath,amssymb,amsfonts}
\usepackage[ruled,linesnumbered]{algorithm2e}
\usepackage{algorithmic}
\usepackage{graphicx}
\usepackage{subfigure}
\usepackage{epstopdf}
\usepackage{textcomp}
\usepackage{xcolor}
\usepackage{multicol}
\def\BibTeX{{\rm B\kern-.05em{\sc i\kern-.025em b}\kern-.08em
    T\kern-.1667em\lower.7ex\hbox{E}\kern-.125emX}}

\ifodd 1
\newcommand{\com}[1]{\textbf{\color{blue} (COMMENT: #1)}} 
\else

\newcommand{\com}[1]{}
\fi

\begin{document}

\title{Placement Optimization of Aerial Base Stations with Deep Reinforcement Learning
}
\author{Jin~Qiu, Jiangbin~Lyu,~\textit{Member,~IEEE},
        and~Liqun~Fu,~\textit{Senior Member,~IEEE}
 \thanks{The authors are with School of Informatics, Xiamen University, China 361005 (email: qeauty@stu.xmu.edu.cn; \{ljb, liqun\}@xmu.edu.cn). \textit{Corresponding Author: Jiangbin Lyu.}}%
}


\maketitle

\begin{abstract}
Unmanned aerial vehicles (UAVs) can be utilized as aerial base stations (ABSs) to assist terrestrial infrastructure for keeping wireless connectivity in various emergency scenarios. To maximize the coverage rate of $N$ ground users (GUs) by jointly placing multiple ABSs with limited coverage range is known to be a NP-hard problem with exponential complexity in $N$. The problem is further complicated when the coverage range becomes irregular due to site-specific blockage (e.g., buildings) on the air-ground channel in the 3-dimensional (3D) space. To tackle this challenging problem, this paper applies the Deep Reinforcement Learning (DRL) method by 1) representing the state by a \textit{coverage bitmap} to capture the spatial correlation of GUs/ABSs, whose dimension and associated neural network complexity is invariant with arbitrarily large $N$; and 2) designing the action and reward for the DRL agent to effectively learn from the dynamic interactions with the complicated propagation environment represented by a 3D Terrain Map. Specifically, a novel two-level design approach is proposed, consisting of a preliminary design based on the dominant line-of-sight (LoS) channel model, and an advanced design to further refine the ABS positions based on site-specific LoS/non-LoS channel states. The double deep Q-network (DQN) with Prioritized Experience Replay (Prioritized Replay DDQN) algorithm is applied to train the policy of multi-ABS placement decision. Numerical results show that the proposed approach significantly improves the coverage rate in complex environment, compared to the benchmark DQN and K-means algorithms.
\end{abstract}

\section{Introduction}
%
%
%
%
%
%

With their high mobility and reducing cost, unmanned aerial vehicles (UAVs) have attracted increasing interests in military and civilian domains in recent years. In particular, integrating UAVs into cellular networks as aerial base stations (ABSs) to assist terrestrial communication infrastructure in various emergency scenarios such as battlefields, disaster scenes and hotspot events, has been regarded as an important and promising technology \cite{zeng2016wireless}.

One of the key problems in UAV-aided communication is to find applicable placement of ABSs aiming to achieve maximum coverage of ground users (GUs).
To maximize the coverage rate of $N$ GUs by jointly placing multiple ABSs with limited coverage range is known to be a NP-hard problem with exponential complexity in $N$\cite{Lyu2016Placement}.
However, it has still spurred enthusiasm of many researchers in this theme\cite{Lyu2016Placement,7461487,LAPlosProbability,PlacementCirclePacking,8760267,8644345,Liu2018Energy}.
The authors in \cite{Lyu2016Placement} propose a spiral algorithm to place ABSs along a spiral path to cover all GUs with the minimum number of ABSs, which reduces the complexity to polynomial-time.
A heuristic K-means clustering algorithm is applied in \cite{7461487}, which finds suitable ABS locations to serve the partitioned GUs.
In terms of maximizing the coverage area, the authors in \cite{LAPlosProbability} optimize the altitude of a single ABS based on the probabilistic line-of-sight (LoS) channel model, while circle packing theory is used in \cite{PlacementCirclePacking} to maximize the total coverage area of multiple ABSs.
On the other hand, controlling ABS movement to cover moving users is another challenging task \cite{8760267}\cite{8644345}, for which \cite{8760267} applies a majority rule to control the direction and distance of UAV displacement towards the cell with the highest user density, while \cite{8644345} uses the K-means algorithm to partition GUs into clusters, and further applies the Q-learning algorithm for ABS movement.
In addition, in terms of ABS coverage and energy consumption trade-off, a Deep Reinforcement Learning (DRL)-based approach is proposed in \cite{Liu2018Energy} to achieve energy-efficient and fair communication coverage.

\begin{figure}[t]
  \centering
  \hspace{-10pt}
  \subfigure[]{
    \label{fig:subfig:a} 
    \includegraphics[width=0.45\linewidth, trim=0 20 0 20,clip]{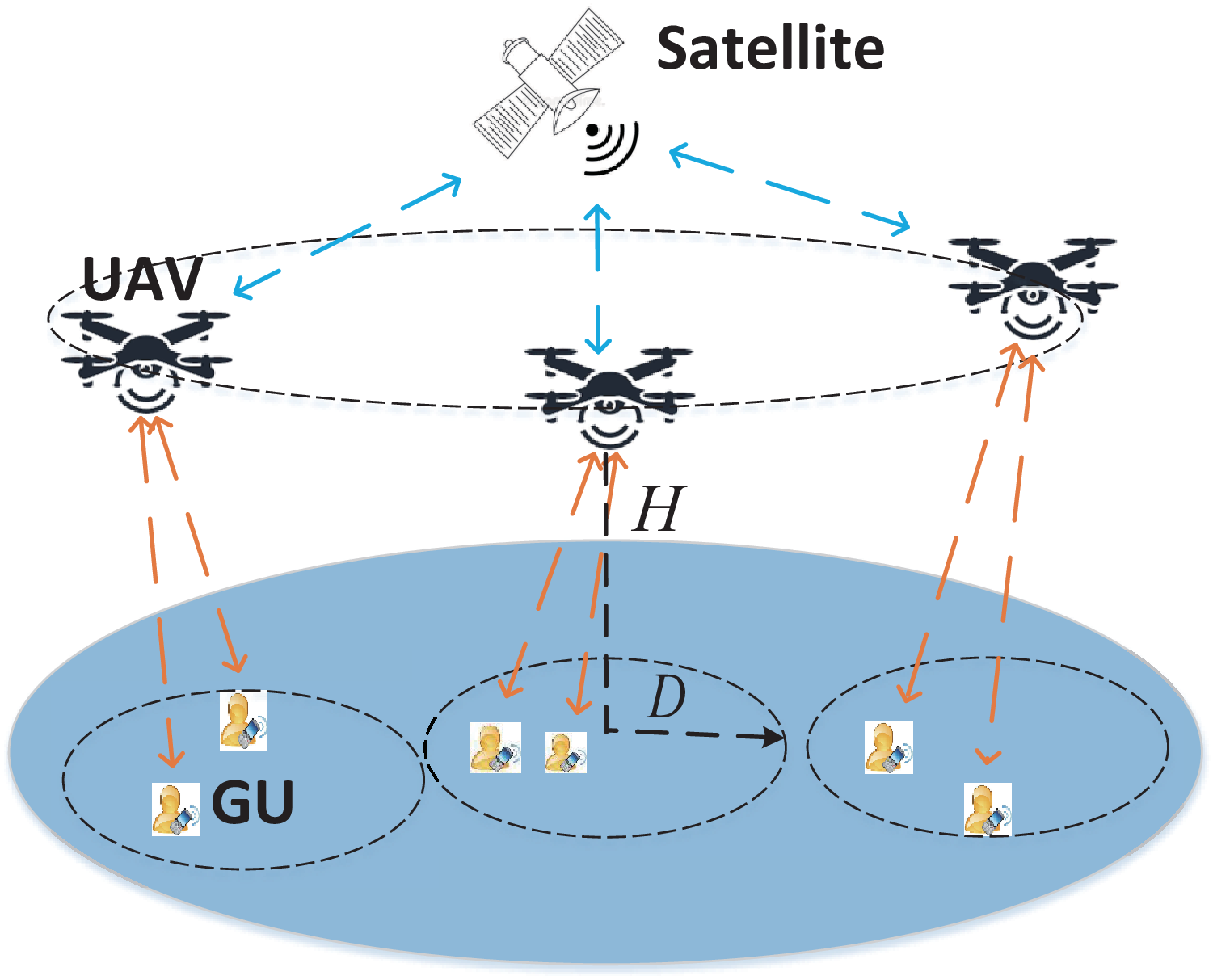}}
  \subfigure[]{
    \label{fig:subfig:b} 
    \includegraphics[width=0.45\linewidth, trim=0 0 0 0,clip]{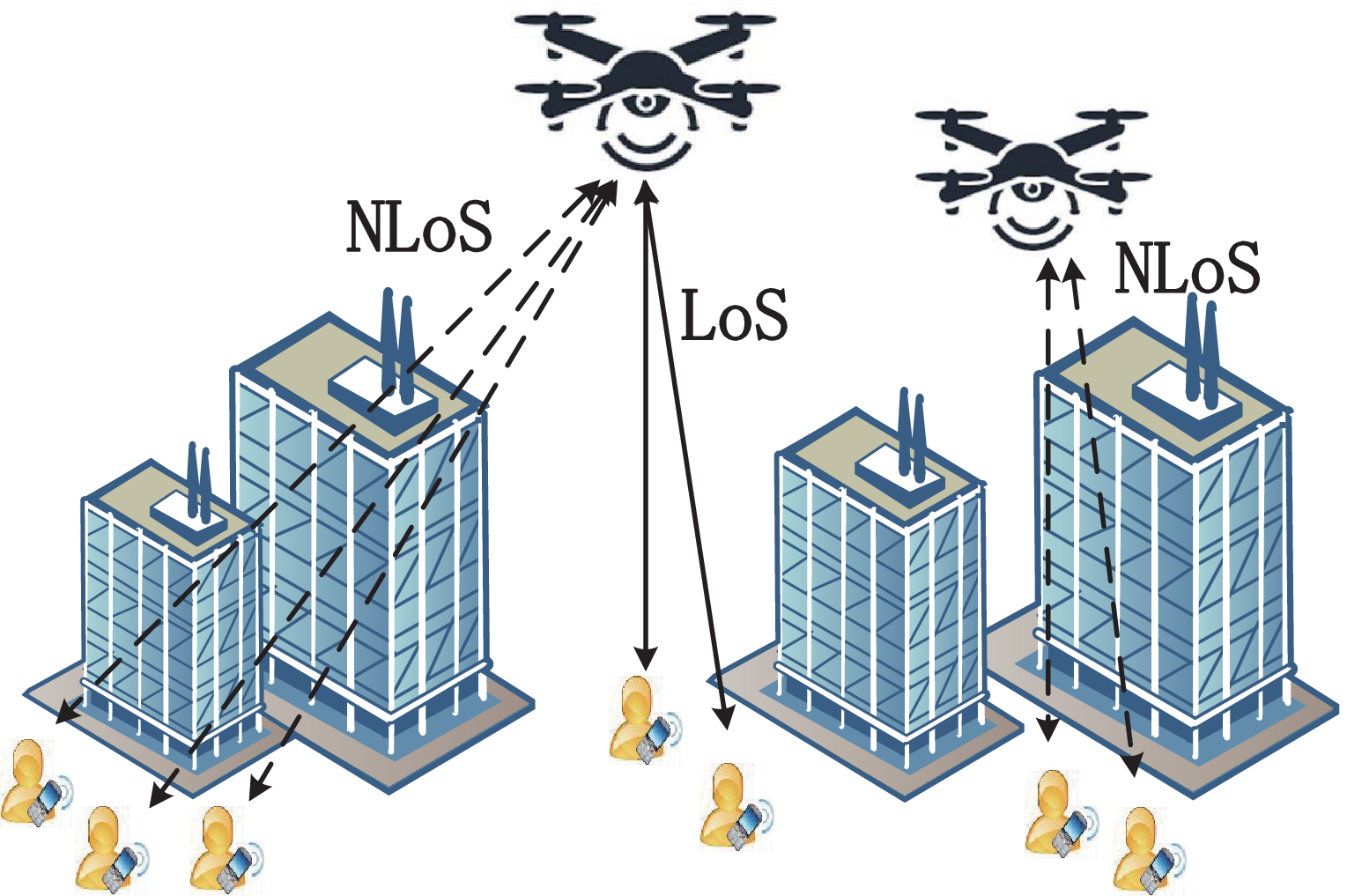}}
  \caption{Placement optimization of ABSs by (a) preliminary design based on dominant-LoS channel and (b) advanced design based on 3D Terrain Map.\vspace{-2ex}}
  \label{fig:subfig} 
\end{figure}

In the aforementioned works, the ABS-GU communication range is determined by a certain signal-to-noise ratio (SNR) threshold, by assuming the air-ground channel to follow the dominant-LoS or probabilistic LoS model\cite{LAPlosProbability}, thus resulting in uniform coverage range (or disk coverage area).
However, the above channel models fail to capture the fine-grained structure of the LoS or non-LoS (NLoS) propagation at specific ABS and GU locations, which in turn critically affects the ABS-GU channel and hence the coverage performance of practical ABS deployment.
For example, with slight change of its position, an ABS might transit from LoS to NLoS propagation to the GU due to building edges.
Such site-specific LoS/NLoS propagation has been exploited in \cite{LOSmap} to find the optimal UAV-relay position for a given pair of ground BS and user.
In the paradigm of cellular-connected UAV\cite{CellularConnectedUAVmag}\cite{LyuIoTJ3d},
the LoS/NLoS channel state can be estimated by the UAV on-site\cite{zeng2019path}, or obtained from a given 3D Terrain Map\cite{zhang2019radio}, based on which the trajectory of the aerial (UAV) user can be optimized to avoid cellular coverage holes and/or minimize flying distance.
However, these works \cite{LOSmap}\cite{zeng2019path}\cite{zhang2019radio} consider a single UAV as an aerial relay or user, with different setup and objective from those in our considered multi-ABS/multi-GU coverage problem.

Due to the site-specific propagation in the 3-dimensional (3D) space,
the LoS/NLoS channel states for all pairs of possible ABS-GU locations in a given environment ensemble an enormous and irregular state space, which cannot be readily handled by conventional optimization methods to achieve maximum coverage rate, especially when the number of ABSs/GUs is large.
To tackle this challenging problem, we
propose a novel two-level design approach, consisting of a preliminary design based on the dominant-LoS channel model, and an advanced design to further refine the ABS positions based on the 3D Terrain Map, as shown in Fig. \ref{fig:subfig}.
For each design, we apply the state-of-the-art double deep Q-network (DQN) with Prioritized Experience Replay (Prioritized Replay DDQN) method, with tailored incorporation of the domain knowledge, by 1) representing the state by a \textit{coverage bitmap} to capture the spatial correlation of GU/ABS locations, which is well fit as the input of the underlying deep neural network (DNN), whose dimension and associated DNN complexity is invariant with arbitrarily large $N$; and 2) designing the action and reward for the DRL agent to effectively learn from the dynamic interactions with the complicated propagation environments.
Numerical results show that the proposed approach significantly improves the coverage rate of GUs compared to the benchmark DQN and K-means algorithms.
Moreover, the advanced design further improves the accuracy of GU coverage over the preliminary design, by exploiting the fine-grained structure of the complex propagation environment.

\section{System Model}
Consider a UAV-aided communication system with $M$ UAV-mounted ABSs to serve a group of $N$ GUs with given locations denoted by $\bold{w}_n\in\mathbb{R}^2$, $n\in\mathcal{N}\triangleq \{1,\cdots,N\}$. Consider downlink communication from ABSs to GUs, while the proposed approach can be similarly applied to uplink communication. Assume that the UAVs fly at a fixed altitude $H$ meters (m), with horizontal locations denoted by $\bold{u}_m\in\mathbb{R}^2$, $m\in\mathcal{M}\triangleq\{1,\cdots,M\}$.
To focus on the coverage performance, we assume for simplicity that the available spectrum is equally divided into $N$ orthogonal channels, each allocated to one GU, and thus there is no intra or inter-cell interference.
Next, we introduce the channel models and coverage criteria for ABS-GU communications.

\subsection{Dominant-LoS Channel Model}
Due to the high altitude of the UAV, LoS channel exists with a high probability for practical ABS-GU links\cite{3GPP}.
In the preliminary design without site-specific information, we assume the dominant-LoS channel model for the ABS-GU communication.
As a result,
the channel power gain between ABS $m$ and GU $n$ is given by
\begin{equation}\label{LoSchannel}
{g_{m,n}} \triangleq \frac{{{\beta _0}}}{{d_{m,n}^2 + {H^2}}},
\end{equation}
where $\beta_0=(\frac{4\pi f_c}{c})^{-2}$ denotes the channel power gain at a reference distance of 1 m, with $f_c$ denoting the carrier frequency and $c$ denoting the speed of light; and $d_{m,n}\triangleq\Vert \bold{u}_m-\bold{w}_n \Vert$ denotes the horizontal distance, with $\Vert \cdot \Vert$ denoting the Euclidean norm.

Assume that each ABS or GU is equipped with a single omni-directional antenna with unit gain.
Assume that each ABS transmits with power $P$ Watt (W) to its served GU, whose receiver noise power is denoted by $\sigma^2$ W.
The SNR received by GU $n$ from ABS $m$ is then given by
\begin{equation}\label{gamma}
\gamma_{m,n}\triangleq g_{m,n}P /\sigma^2.
\end{equation}
A GU is said to be \textit{covered} by an ABS, if the received SNR is not smaller than a certain threshold $\bar\gamma$, which corresponds to $g_{m,n}\geq \bar\gamma \sigma^2/P\triangleq \bar g$, with $\bar g$ denoting the corresponding threshold of channel power.
For the LoS channel model in \eqref{LoSchannel}, $\bar g$ further corresponds to a distance threshold $D$ (also known as \textit{coverage range}) such that $d_{m,n}\leq D$, which is given by
\begin{equation}
D \triangleq \sqrt {{\beta _0}/\bar g - {H^2}},
\end{equation}
as illustrated in Fig. \ref{fig:subfig}(a).
Finally, denote $C_n$ as the \textit{coverage indicator} for GU $n$, which is given by
\begin{equation}\label{CnDisk}
C_n\triangleq\begin{cases}
1, & \textrm{if} \ \min\limits_{m \in \mathcal{M}}d_{m,n}\leq D, \\
0, & \textrm{otherwise.}
\end{cases}
\end{equation}

\subsection{Site-Specific LoS/NLoS Channel Model}

Despite the high LoS probability, the air-ground channel could be occasionally obstructed by obstacles, resulting in NLoS propagation.
To investigate the large-scale coverage performance, we assume that the small-scale fading effect is averaged out, and thus focus on the dominant LoS and NLoS path-loss components, as in \cite{LAPlosProbability}.
In the case when the 3D Terrain Map for a specific environment can be obtained, e.g., from geographic information system (GIS), we can extract the LoS/NLoS information for any pair of ABS and GU locations, as shown in Fig. \ref{fig:subfig}(b).
Therefore, the channel power gain between ABS $m$ and GU $n$ can be expressed as
\begin{align}\label{probLOS}
g_{m,n}\triangleq
\begin{cases}
g_{\textrm{L}}(\bold u_m,\bold w_n), & \quad \textrm{without obstacles in between;}\\
g_{\textrm{NL}}(\bold u_m,\bold w_n), & \quad \textrm{otherwise,}
\end{cases}
\end{align}%
where $g_{\textrm{L}}$ and $g_{\textrm{NL}}$ denote the channel power gains of the LoS and NLoS channels, respectively, whose specific function forms can be referred to the empirical formula in \cite{3GPP}.
In this case, the coverage indicator for GU $n$ is given by
\begin{equation}\label{CnMap}
C_n\triangleq\begin{cases}
1, & \textrm{if} \ \max\limits_{m \in \mathcal{M}}g_{m,n}\geq \bar g, \\
0, & \textrm{otherwise.}
\end{cases}
\end{equation}

\section{Problem Formulation}
Define the \textit{coverage rate} of all GUs as the ratio of GUs covered by at least one of the ABSs, i.e., $\varphi\triangleq \frac{1}{N}\sum_{n\in\mathcal{N}} C_n $.
We formulate the placement optimization problem to maximize the coverage rate of GUs with $M$ ABSs, given by
\begin{align}
\textrm{(P1):}&\max\limits_{\bold u_m, m\in\mathcal{M}}\ {\rm{ }}\varphi\triangleq \frac{1}{N}\sum_{n\in\mathcal{N}} C_n, \notag \\
&\textrm{s.t.}\quad C_n \textrm{ given by \eqref{CnDisk} or \eqref{CnMap}, for $n\in\mathcal{N}$}. \notag
\end{align}

For the preliminary design with $C_n$ given by \eqref{CnDisk}, (P1) is a non-convex optimization problem due to the non-convex constraint of $\min\limits_{m \in \mathcal{M}}d_{m,n}\leq D$. In fact, it is shown to be a NP-hard problem\cite{Lyu2016Placement} in general, with exponential complexity in $N$.
The problem is further complicated in the advanced design with $C_n$ given by \eqref{CnMap}, where the LoS/NLoS channel states for all pairs of possible ABS-GU locations ensemble an enormous and irregular state space, which cannot be readily handled by conventional optimization methods, especially when the number of ABSs/GUs is large.

To tackle this challenging problem, we apply the state-of-the-art Prioritized Replay DDQN method, with tailored considerations of the domain knowledge, by 1) representing the state by a coverage bitmap to capture the spatial correlation of GU/ABS locations, which is well fit as the input of the underlying DNN, whose dimension and associated DNN complexity is invariant with arbitrarily large $N$; and 2) designing the action and reward for the DRL agent to effectively learn from the dynamic interactions with the complicated propagation environments.
The DRL framework possesses general intelligence to solve complex problems, which is able to handle with the large and complicated state space involved in the problem (P1) and solve it effectively.

\section{Placement Optimization with DRL}
\subsection{DRL Algorithm}
This subsection gives a brief introduction on the DRL algorithm before presenting the proposed design.
In general, DRL is the combination of DNN and RL. Specially, RL refers to the process in which an agent interacts with the environment and makes a series of decisions by using Markov Decision Process (MDP) \cite{sutton2018reinforcement}. At each time step $t$, the agent observes state ${s_t}$,
executes action ${a_t}$, and then receives instant reward ${r_t}$, and transits to the next state
${s_{t+1}}$, which forms a sequence $\left\langle {{s_t},{a_t},{r_t},{s_{t + 1}}} \right\rangle $
of MDP.
Define the return ${G_t}$ as the sum of discounted rewards, given by
\begin{equation}
{G_t} \triangleq {r_t} + \beta {r_{t + 1}} + {\beta ^2}{r_{t + 2}} +  \cdots  = \sum\limits_{k = 0}^{\infty} {{\beta ^k}{r_{t + k }}},
\end{equation}
where $0<\beta <1 $ denotes the discount factor.
Define \textit{policy} $\pi$ as the state-to-action mapping.
The agent optimizes policy $\pi$ in order to maximize the action-value function ${Q}$ defined as
\begin{equation}
{Q}\left( {s,a|\pi} \right) \triangleq \mathbb{E}_{\pi}[{G_t}|{s_t} = s,{a_t} = a],
\end{equation}
which is the expectation of the return ${G_t}$ at the current state $s$ and action $a$ under policy $\pi$.

\indent However, RL can only handle problems with small state space and action space, which is inappropriate for our problem. To this end, we use DNN as the approximator of the $Q$ function in Q-learning \cite{sutton2018reinforcement}, which constitutes a commonly-used DRL framework known as DQN. In particular, the algorithm applies \textit{experience replay} to sample data offline, and \textit{target network mechanism} that modifies action-value $Q$ towards target values to
improve algorithm convergence. The DQN is trained to minimize the loss function defined as
\begin{equation}\label{lossFunction}
L({\theta }) \triangleq \mathbb{E}\big[\big({{y_t} -{Q({s_t},{a_t}|{\theta })}}\big)^2\big],\\
\end{equation}
where the vector $\theta$ represents the DQN weights that determine the policy $\pi$, and ${y_t}$ is the target function given by
\begin{equation}\label{targetf}
{y_t} \triangleq r_t + \beta \mathop {\max }\limits_{{a}} {Q}({s_{t + 1}},{a}|{\theta_\textrm{target}}),
\end{equation}
where $\theta_\textrm{target}$ is copied from $\theta$ every fixed number of steps.

Despite the efficiency of DQN, it still has some critical limitations: 1) Overestimations have been attributed to the greedy algorithm used by the target function, which negatively affects the performance of policy; 2) Uniform samples have been applied in experience replay rather than weighted samples based on significance, which may lead to divergence in target with large state space. In order to overcome the above limitations, we apply the Prioritized Replay DDQN to address the problem, which mainly improves in two aspects. First, \eqref{targetf} is adapted as\\
\begin{equation}\label{ytDQ}
y_t^{DQ} \triangleq r_t + \beta Q({s_{t + 1}},\mathop {\arg}\mathop {\max}\limits_{a} Q({s_{t + 1}},a;{\theta})|{\theta_\textrm{target}}),
\end{equation}
which untangles the selection and evaluation respectively in Q-learning to avoid overestimation \cite{van2016deep}. Second, experiences are replayed with prioritized sampling using the sum-tree structure \cite{schaul2015prioritized}, which is updated efficiently.\\

\subsection{Preliminary Design Based on LoS Channel Model}
In this subsection, we aim to design the ABS placement to maximize the coverage rate of GUs under the dominant-LoS channel model using the DRL framework.
To achieve fast convergence, we apply the DRL algorithm phase by phase. In each phase, we set a target coverage rate $\bar\varphi$ and train the underlying DNN towards achieving $\bar\varphi$.
The target coverage rate $\bar\varphi$ is then gradually increased until it can no longer be achieved, by which a suboptimal solution to (P1) is obtained.
Specifically, for each phase, we cast the placement problem into a MDP, and define the state-action-reward tuple $\left\langle s,a,r\right\rangle $ with our domain knowledge as follows.

1) State ${s}$: Normally, state represents the input of DNN. A straightforward choice of the state is the profile of all GU and ABS locations, whose dimension and associated complexity increases with $N$ and $M$. Moreover, the neural network is not sensitive to the scalar-type location variables without proper quantification.
Therefore, a more suitable form of state representation is desired, for which we propose the \textit{coverage bitmap}.
Specifically, we equally partition the considered (rectangular) ground area $\mathcal{G}$ into $K$-by-$K$ grid regions $\mathcal{G}_{ij}$, $i, j\in\mathcal{K}\triangleq \{1,\cdots,K\}$.
Denote the number of covered GUs in region $\mathcal{G}_{ij}$ as
\begin{equation}\label{fij}
f_{ij}\triangleq \sum_{\bold w_n\in\mathcal{G}_{ij}} C_n.
\end{equation}
As a result, we choose the state $s=F\triangleq {[{f_{ij}}]_{K \times K}}$, where the matrix $F$ is in the form of a 2D bitmap, which effectively captures the spatial correlation of the GU and ABS locations in terms of the number $f_{ij}$ of covered GUs in each grid, and thus termed \textit{coverage bitmap}.
Moreover, the bitmap data structure is well fit as the input type of the state-of-the-art DNN (more specifically, the convolutional neural network (CNN)), whose input dimension ($K\times K$) and associated DNN complexity is invariant with arbitrary large $N$, thus circumventing the curse of dimensionality in DNN.
Note that the selection of $K$ still needs to balance between the bitmap resolution and the complexity of DNN. However, a moderately large $K$ would suffice since the detailed spatial correlation of GU and ABS locations are effectively represented (and weighted) by $f_{ij}$ and nested into the state matrix $F$.

2) Action $a$: For simplicity, assume that the action space of each ABS in each time step consists of four moving operations \{up, down, left, right\} with a certain displacement size $\Delta$ m. The overall action $a$ is then an $M$-dimensional vector.

3) Reward ${r}$: In our context of maximizing the coverage rate, the reward $r_t$ in each time step $t$ needs to encourage the state-action pair that brings the current coverage rate $\varphi_t$ closer to the ideal value of 1. Therefore, we choose the negative error function as the reward for the intermediate steps with $\varphi_t<\varphi$.
When the target coverage rate is achieved, i.e., $\varphi_t\geq\varphi$, we set a positive reward $r_t=1$ and terminate the episode. In addition, when ABSs are out-of-border\footnote{We define \textit{out-of-border} as the case with at least two ABSs flying beyond the border, in order not to receive too much negative rewards.}, we set a negative reward $r_t=-1$ to punish such behavior. Thus, the reward function is defined as
\begin{align}\label{R}
r_t \triangleq\begin{cases}
-{\alpha\left(\varphi_t - 1 \right)^2}, & \textrm{if}\ \varphi_t < \bar\varphi,\\
1, & \textrm{if}\ \varphi_t \ge \bar\varphi, \\
-1, & \textrm{if ABSs are out-of-border,}
\end{cases}
\end{align}
where $\alpha$ is a positive constant to scale the reward.

Based on the defined state-action-reward tuple $\left\langle s,a,r \right\rangle $, the proposed ABS placement optimization with Prioritized Replay DDQN is presented in Algorithm \ref{algorithm1}.
The algorithm starts by initializing the parameters (Line 1), followed by $E$ episodes.
Each episode starts by resetting the state (Line 2), followed by $T$ steps.
Each step consists of the \textit{exploration} part (Lines 4$\sim$7) and the \textit{training} part (Lines 8$\sim$17). In the exploration part, the agent interacts with the environment by observing the current state $s_t$, choosing an action $a_t$ based on policy $\pi$, and obtaining the next state $s_{t+1}$ and instantaneous reward $r_t$.
The transition $\{s_t,a_t,r_t,s_{t+1}\}$ is then stored in memory $\mathcal{H}$.
After memory $\mathcal{H}$ is full,
the training part starts by sampling a minibatch\footnote{Minibatch is used such that the model updates are fast (as opposed to processing the whole training data) and not too noisy (as opposed to processing every instance).} of $l$ transitions for the training process to update the weights $\theta$ in order to minimize the weighted loss function, which is given by
\begin{equation}\label{lossf}
L(\theta) \triangleq \frac{1}{l}\sum\limits_{j = 1}^l \omega _j \delta_j^2=\frac{1}{l}\sum\limits_{j = 1}^l {{\omega _j}{{\big({y_j} - Q({s_j},{a_j}|{\theta})\big)}^2}}.
\end{equation}
In \eqref{lossf}, ${\delta_j} \triangleq y_j - Q({s_j},{a_j}|{\theta})$ denotes the Temporal-Difference (TD)-error, $y_j$ is the target function given by \eqref{ytDQ},
and ${\omega _j}$ denotes the importance-sampling weight\cite{schaul2015prioritized} used to correct the bias, which is given by\\
\begin{equation} \label{importants}
{\omega _j} \triangleq \frac{{{{\big(|\mathcal{H}| \cdot P(j)\big)}^{ - \nu }}}}{{{{\max }_i}{\omega _i}}},
\end{equation}
where $|\mathcal{H}|$ is the memory size and $\nu$ is a positive constant.
In the training process, transition $j$ is sampled based on the probability given by
\begin{equation}\label{probability}
P(j) \triangleq \frac{{p_j^{\mu} }}{{\sum\nolimits_{i=1}^l {p_i^{\mu} } }},
\end{equation}
where $p_j$ denotes the priority of transition $j$, and $\mu>0$ denotes the degree of priority.
The priority $p_j$ is initialized to be 1 for all samples before memory $\mathcal{H}$ is full, so that they all stand a chance to be sampled. After memory $\mathcal{H}$ is full, we set $p_j=|\delta_j|$ to attribute a higher priority to the transition with larger absolute TD-error (which suggests greater model mismatch).
The episode is terminated if $\mathcal{H}$ is full, and the target coverage rate $\bar\varphi$ is achieved, i.e., $r_t=1$ (Line 18).


\begin{algorithm}[t]\small  \label{algorithm1}
	\caption{ABS placement optimization with Prioritized Replay DDQN}
	{{
	\KwIn{GU locations, initial ABS locations (and 3D Terrain Map), and coverage range $D$ (or channel power threshold $\bar g$).}	
	\KwOut{Final ABS locations and achieved coverage rate.}
	Initialize target coverage rate $\bar\varphi$, replay memory $\mathcal{H} = \emptyset $, $p_1=1$.
	
	\For{$episode: =1,\cdots,E$}
	{
		Initialize the environment and receive an initial state $s_1$.\\
		\For{$step \ t: =1,\cdots,T$}
		{
			${a_t} = \pi ({s_t})$;\\
			Execute $a_t$, and obtain $s_{t+1}$ and $r_t$ in \eqref{R};\\
			Store transition $\{s_t,a_t,r_t,s_{t+1}\}$ in $\mathcal{H}$ with maximal priority
			${p_t} = {\max _{i < t}} \ {p_i}$.\\
			\If{\textrm{$\mathcal{H}$ is full}}
			{
				\For{$j:=1,\cdots,l$}
				{
					Sample transition $j$ based on \eqref{probability};\\
					Compute importance-sampling weight in \eqref{importants};

                    Compute TD-error $\delta_j$;\\

					Update transition priority ${p_j} \leftarrow \left| {{\delta _j}} \right|$.
				}
				Update weights $\theta$ of $Q( \cdot )$ by minimizing the loss function in \eqref{lossf};\\
				Set ${\theta _\textrm{target}} =\theta $ every fixed number of steps.
			}
			Terminate the episode if $r_t=1$ when $\mathcal{H}$ is full.\\
		}	
	}
	Increase the target coverage rate $\bar\varphi$ and repeat Lines 1$\sim$20, until it can no longer be achieved.
}}
\end{algorithm}

Through the Prioritized Replay DDQN (Lines 1$\sim$20), the DNN weights $\theta$ are trained to minimize the loss function, thus fitting the DNN towards achieving the target function in \eqref{ytDQ}, which in turn approximates the maximum action-value $Q$ (or expected sum of discounted rewards), and hence improving the achieved coverage rate.
In particular, the proposed reward function in \eqref{R} encourages the state-action pair that brings the coverage rate closer to 1, and punishes ABS out-of-border behavior.
Together with the proposed state representation by coverage bitmap, we have coherently incorporated the domain knowledge in our problem context into the DRL framework, and thus able to solve the complicated problem (P1) effectively.

\subsection{Advanced Design Based on 3D Terrain Map}\label{SectionIV-C}
\begin{figure}[t]
	\centering\vspace{-2ex}
	\includegraphics[height=5cm,width=7cm]{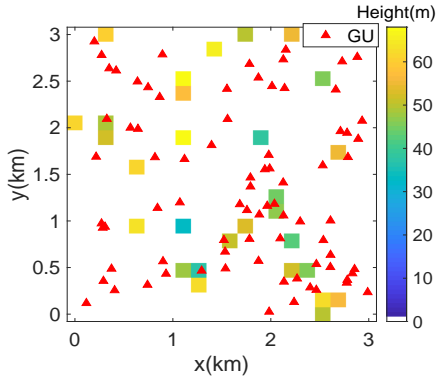}
	\caption{An example of 3D Terrain Map with 80 GUs inside.\vspace{-2ex}}
	\label{Terrain map}
\end{figure}
The preliminary design optimizes the ABS placement under the dominant-LoS channel model. To achieve more accurate coverage in the site of interest, we assume that its 3D Terrain Map is available, based on which we propose the advanced design to further refine the ABS positions tailored to the specific environment.
For illustration, we generate an example of 3D Terrain Map for a square region of side length 3 km, with 30 buildings and 80 GUs randomly located inside, as shown in Fig. \ref{Terrain map}.
The buildings are modeled by cuboids with length and width of 150 m, and random height based on uniform distribution in $[30,70]$ m.

In the advanced design, the definitions of $\left\langle s,a,r\right\rangle $ are similar to those in the preliminary design, despite the calculation of the coverage bitmap.
Specifically, the coverage bitmap $F$ now relies not only on the ABS-GU distance, but also the LoS/NLoS channel state for the specific ABS/GU locations.
To this end, we propose Algorithm \ref{algorithm2} to obtain the coverage bitmap based on the 3D Terrain Map.
Algorithm \ref{algorithm2} begins by resetting the coverage bitmap and coverage indicators (Line 1).
For each GU $n$, we first sort the ABSs by ascending order of the GU-ABS distance (i.e., from near to far), and then check the GU coverage by the sorted ABS order (Lines 2$\sim 10$).
This helps to reduce the computational complexity, since a nearer ABS is more likely to have LoS channel with the GU and hence cover it, thanks to the air-ground channel characteristics\cite{3GPP}.
The obtained coverage indicator $C_n, n\in\mathcal{N}$ is then accumulated for each grid region $\mathcal{G}_{ij}$ to obtain $f_{ij}$ as in \eqref{fij} (Line 11), and hence the coverage bitmap $F={[{f_{ij}}]_{K \times K}}$.

\begin{algorithm}\footnotesize  \label{algorithm2}
	\caption{Obtaining the coverage bitmap based on 3D Terrain Map}
	{
	\KwIn{GU locations, ABS locations, 3D Terrain Map and channel power gain threshold $\bar g$. }
	
	\KwOut{Matrix form of coverage bitmap $F={[{f_{ij}}]_{K \times K}}$.}
	Initialize $F=0$ and $C_n=0, \forall n$.
	
	\For{GU $n=1,\cdots,N$}
	{   Sort the ABSs by ascending order of the GU-ABS distance (i.e., near to far), denoted by the ordered set $\mathcal{M}_\textrm{sorted}$.\\
        \For {ABS $m\in\mathcal{M}_\textrm{sorted}$}
        {
            Calculate $g_{m,n}$ based on \eqref{probLOS};\\
            \If{$g_{m,n}\ge \bar g$}
            { Set $C_n=1$; break.
            }
        }

	}
	Obtain $f_{ij}$ by accumulating $C_n$ for each grid region $\mathcal{G}_{ij}$ as in \eqref{fij}.}
\end{algorithm}

Based on the obtained coverage bitmap, we can then apply the proposed Algorithm \ref{algorithm1} to further refine the ABS placement for the site of interest, by taking the placement result of the preliminary design as initial input.
Note that the preliminary design based on the dominant-LoS channel model captures the main spatial correlation of ABSs/GUs by the \textit{distance-based coverage rule}, while the advanced design based on the 3D Terrain Map exploits the \textit{fine-grained structure} of the air-ground channel, thus able to achieve more accurate coverage in a specific environment.

\section{Numerical Results}
In this section, we present numerical results on the coverage performance of the proposed ABS placement design.
We adopt the example of 3D Terrain Map and GU locations in Fig. \ref{Terrain map}.
The following parameters are used if not mentioned otherwise: $M=10$, $N=80$, $H=90$ m, $D=0.5$ km (corresponding to $\bar g=-93$ dB), $f_c=2$ GHz, $c=3\times 10^8$ m/s, $K=20$, $\Delta=10$ m, $l=64$, $\mu=0.6$, $\nu=0.4$, $T=100$ and $|\mathcal{H}|=40000$.
\begin{figure*}[t]
\centering
\hspace{-6ex}
	\subfigure[]{
		\includegraphics[width=0.75\columnwidth, height=0.55\columnwidth,trim=5 0 5 0,clip]{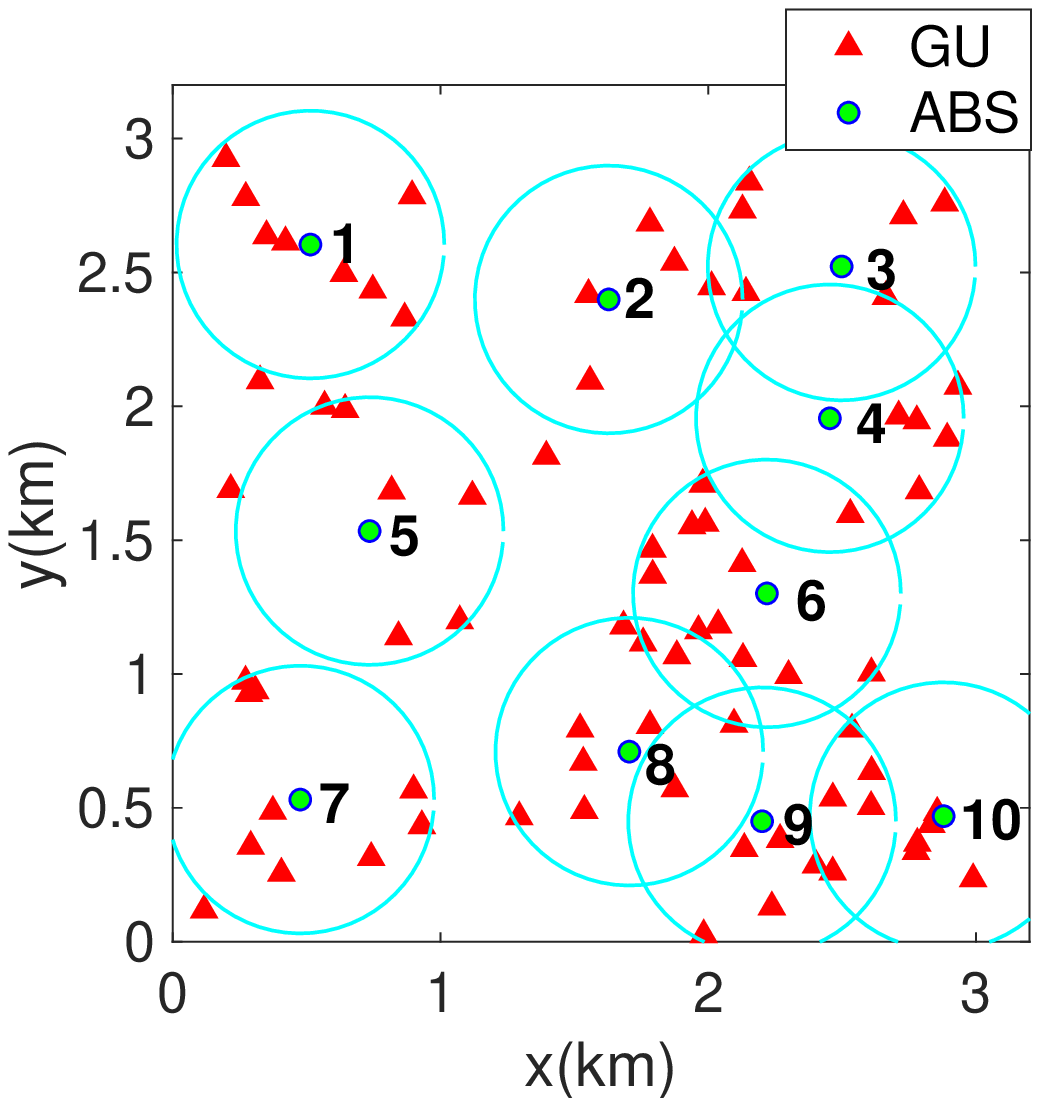}
		\label{fig3a}
	}
	\quad \hspace{-11ex}
	\subfigure[]{
		\includegraphics[width=0.75\columnwidth, height=0.55\columnwidth,trim=5 0 0 0,clip]{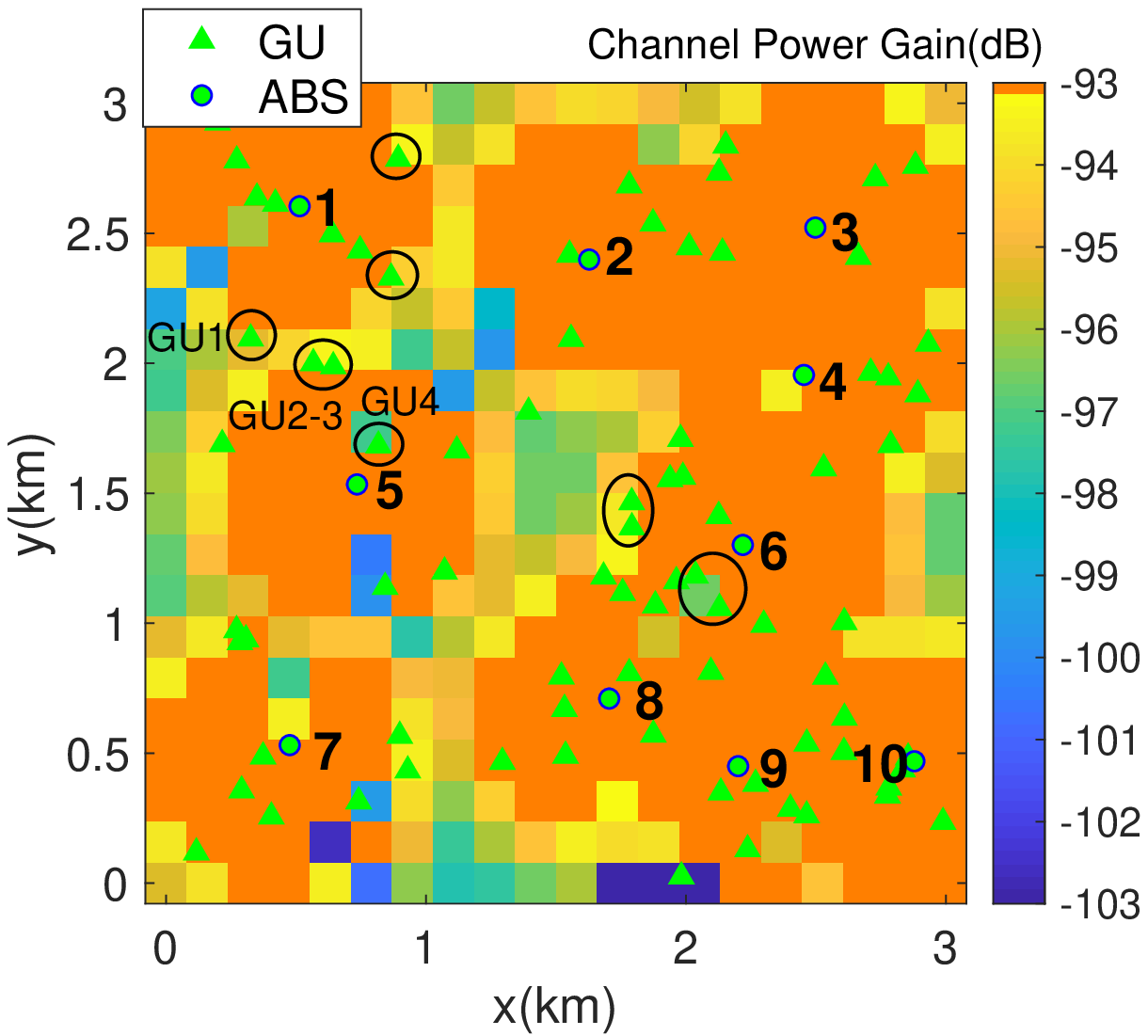}
		\label{fig3b}
	}
	\quad \hspace{-10ex}
	\subfigure[]{
		\includegraphics[width=0.75\columnwidth, height=0.55\columnwidth,trim=0 5 0 0]{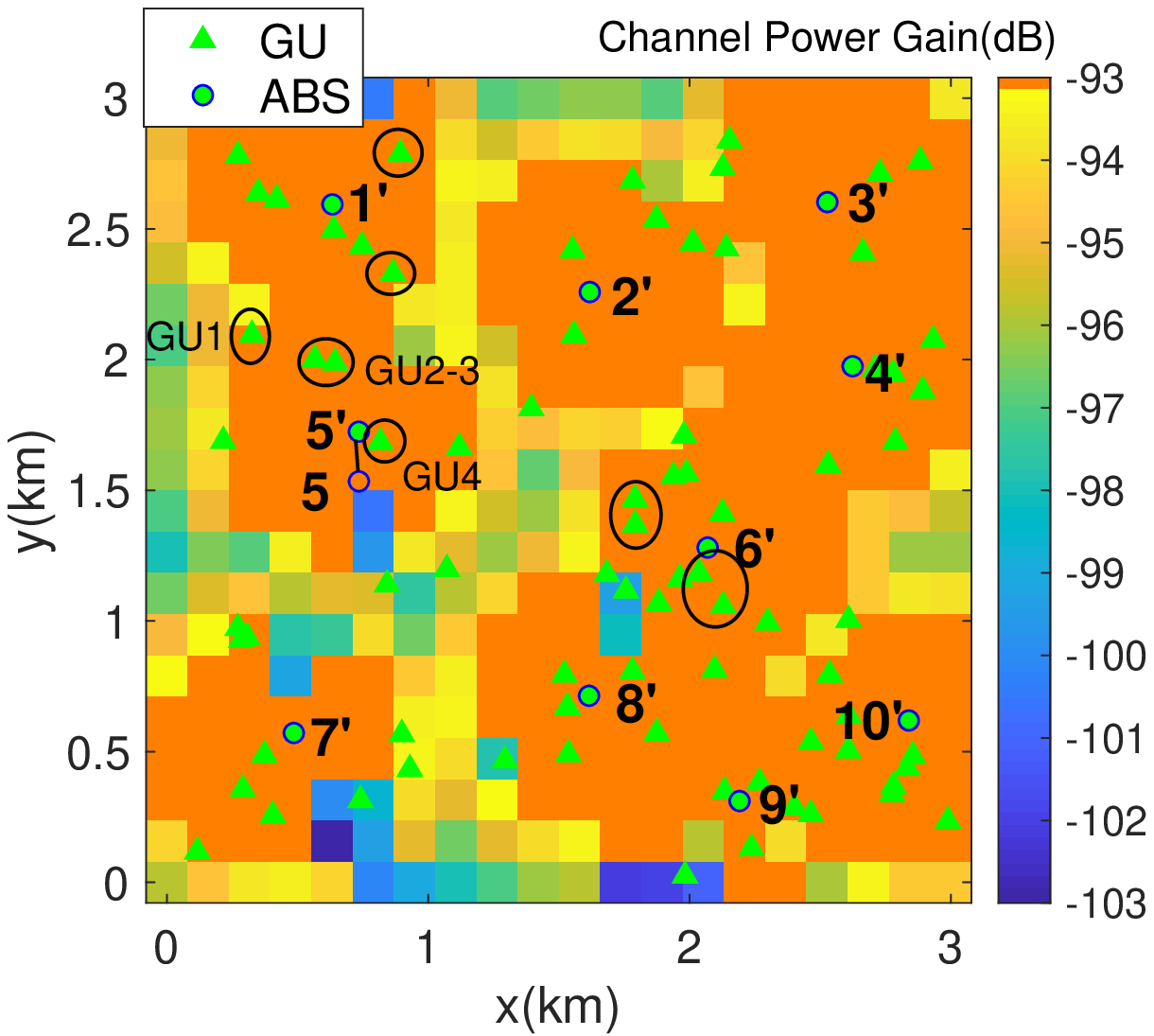}
		\label{fig3c}
	}
\caption{ABS placement and GU coverage by (a) preliminary design with uniform coverage range; (b) preliminary design result applied in the 3D Terrain Map; and (c) advanced design based on the 3D Terrain Map.\vspace{-2ex}}
\label{fig3}
\end{figure*}
\begin{figure}[t]
\centering\vspace{-2ex}
   \includegraphics[width=1\linewidth, trim=0 0 0 0,clip]{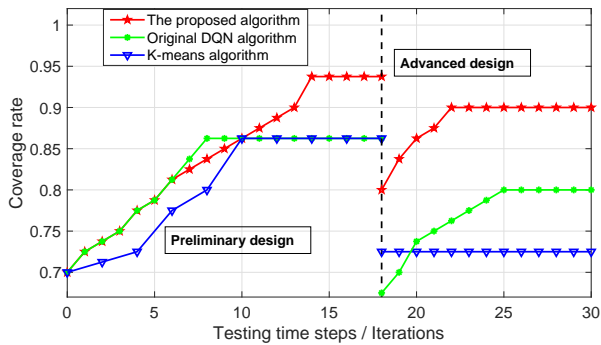}
\caption{Coverage rate achieved by the different algorithms.\vspace{-2ex}}\label{fig4}
\end{figure}


In the preliminary design, the DRL model is trained for 900 episodes from an initial random state, where the obtained placement result is used to set the initial state in the advanced design, which is further trained for 1600 episodes.
The final ABS placement and GU coverage result by the preliminary design with uniform coverage range is shown in Fig. \ref{fig3}(a), where 93.75\% of GUs are covered.
However, when the above result is applied in the considered 3D Terrain Map, the coverage rate drops sharply to 80\%, as shown in Fig. \ref{fig3}(b), due to the NLoS path-loss caused by site-specific blockage in the 3D space.
Fortunately, we further apply the advanced design based on the 3D Terrain Map, which raises the coverage rate back to 90\%, as shown in Fig. \ref{fig3}(c).
Note that in general, the achievable coverage rate based on the 3D Terrain Map is not greater than that based on the dominant-LoS channel model, since the additional signal blockage by obstacles results in NLoS path-loss and hence overall lower coverage rate.
On the other hand, by comparing Fig. \ref{fig3}(b) and Fig. \ref{fig3}(c), it can be seen that some of the GUs (circled in the figures) originally not covered by the preliminary design result, are now covered in the advanced design with slight change of the ABS locations.
For example, GUs 1$\sim$4, originally not covered in the preliminary design, are now covered in the advanced design with slight movement of ABS 5.
This thus demonstrates the benefit of considering site-specific LoS/NLoS channel for the site of interest, and the advantage of our proposed DRL-based ABS placement design in achieving higher coverage rate.

Finally, the coverage rates achieved by the proposed algorithm and the benchmark DQN and K-means algorithms are plotted in Fig. \ref{fig4}, respectively.
It can be seen that the proposed algorithm achieves higher coverage rate in both the preliminary design and advanced design.
On the other hand, in all three algorithms, the achieved coverage rate under the LoS channel model drops when the placement results are applied in the 3D Terrain Map, due to the introduction of additional NLoS path-loss.
Note that the K-means algorithm is distance-based and only applicable for the scenario with uniform coverage range.
In comparison, our proposed algorithm with Prioritized Replay DDQN adapts well in the complex environment, and also outperforms the basic DQN algorithm.

\section{Conclusions}
This paper investigates the placement optimization of multiple ABSs to maximize the coverage rate of GUs under the dominant-LoS channel model first, and further the site-specific LoS/NLoS model.
The problem is NP-hard in general and further complicated by the complex propagation environment.
We tackle this challenging problem using the DRL method by proposing the coverage bitmap as the state representation, which captures the spatial correlation of GUs/ABSs, and is well fit as the input of DNN with fixed dimension and associated complexity.
Moreover, with our proposed action and reward, the DRL agent learns well from the dynamic interactions with the environment using the Prioritized Replay DDQN.
Numerical results show that our proposed design significantly improves the coverage rate compared to benchmark DQN and K-means algorithms.
Our next plan is to extend the current framework to the scenario with moving GUs.
\bibliographystyle{ieeetr}
\bibliography{ref}

\end{document}